\documentclass[preprint,aps,prc,showpacs,showkeys]{revtex4}

\usepackage{graphicx}
\usepackage{bm}

\begin{document}

\title{ Off-shell pions \\ 
        in Boltzmann-Uehling-Uhlenbeck transport theory     
        \footnote{Supported by BMBF and GSI Darmstadt} }

\author{ A.B. Larionov 
         \footnote{On leave from RRC "I.V. Kurchatov Institute", 
                   123182 Moscow, Russia}
         and U. Mosel }

\affiliation{ Institut f\"ur Theoretische Physik, Universit\"at Giessen,
          D-35392 Giessen, Germany }

\date{\today}

\begin{abstract}
Due to large $\pi N \rightarrow \Delta$ cross section a pion acquires
a large width in nuclear matter. Therefore, a $\Delta$-resonance decaying in
the nuclear medium can produce far off-shell pions, which will gradually
lose their virtuality while propagating to the vacuum. To describe such
off-shell dynamics, we implemented a pion spectral function for finite
nuclear matter density -- calculated on the basis of the $\Delta$-hole model
-- in the BUU transport approach. The pion off-shell dynamics leads to an 
enhancement of the low-$p_t$ pion yields in the collisions of Au+Au at 
1 A GeV.
\end{abstract} 



\maketitle 

\section{ Introduction }

Pions are abundantly produced in heavy-ion collisions at beam 
energies of 1 A GeV and above. The main pion production channel
is via the resonance excitation in nucleon-nucleon collisions:
$NN \rightarrow NR$, $R \rightarrow N\pi$, where $R$ stands for 
$\Delta(1232)$ or higher baryonic resonances. As well established
in nuclear transport theory \cite{Eh93,Bass95}, pions propagating
through nuclear matter experience the chain of absorption-production 
processes $\pi_1N_1 \rightarrow R_1 \rightarrow \pi_2N_2 \rightarrow R_2 
\rightarrow \pi_3N_3$ etc due to the large resonance $\pi N$ scattering
cross section. In other words, a pion with momentum $\sim 0.3$ GeV/c
acquires a width in nuclear matter. The pion spectral function 
in nuclear matter is, therefore, different from the one in vacuum.
Actual calculations within the (generalized) $\Delta$-hole model
\cite{Mig74,BW75,EW,HU94,RW94} indeed have shown that the pion spectral 
function is quite broad.

The off-shell nucleon effects in heavy particle (N$^*$(1535)) production
have been studied in \cite{BD96}, where it has been shown, that the 
ground state correlations in colliding nuclei enhance the near-threshold
production substantially. Beyond the ground state correlations, the
influence of the off-shell nucleon propagation on heavy particle production
has been found to be unimportant in \cite{BD96}, while recent BUU 
calculations \cite{EM99,CaJu00} have demonstrated an appreciable 
enhancement of the light meson ($\pi$,K$^+$) production by off-shell 
nucleons.  

The transport simulations with in-medium pion and $\Delta$ properties 
have been recently performed by Helgesson and Randrup in a series
of works (see \cite{Helg98} and Refs. therein) employing the quasiparticle
approximation for pions. The soft pion enhancement due to the in-medium 
pion modifications has been observed in \cite{Helg98} in agreement with 
BUU studies in \cite{Xiong93,Fuchs97}. As has been pointed out in 
\cite{Helg98}, the overall effect of the in-medium modifications on the pion 
observables is small due to the fact, that the observed pions are emitted
from the surface region of the system, where the in-medium modifications
are not important. However, by multistep processes, pions influence other
particles, e.g. kaons. Thus, the in-medium pion modifications could manifest
themselves in the changes of the yields of the other particles.  

The present work is the first attempt to include the pion spectral
function into the BUU transport theory. The structure of the paper
is as follows. In Sect. II we derive the pion spectral function in
nuclear matter at finite temperature on the basis of the $\Delta$-hole
model. Sect. III contains the equation of motion for off-shell pion
propagation. In Sect. IV we present the results of the BUU calculations
for the pion production in Au+Au collisions at 1 A GeV.
The summary of the work is given in Sect. V.

\section{ The pion spectral function in nuclear matter }

In the $\Delta$-hole model the pion polarization function $\Pi(k)$
is given by the loop shown in Fig.~\ref{fig:dhole}, where we neglected
the exchange $u$-diagram.
For the evaluation of $\Pi(k)$ we use a simple Lagrangian of the form:
\begin{equation}
      {\cal L}_{\pi N \Delta} = 
      {f_{\pi N \Delta} \over m_\pi} \bar\psi_\Delta^\mu {\bf T} \psi_N
      \partial_\mu \mbox{\boldmath ${\mathbf \pi}$ \unboldmath} + h.c.~,
                                                           \label{Lagr} 
\end{equation}
where ${\bf T}$ is the isospin transition operator ($1/2\rightarrow3/2$)
\cite{BW75}, $\psi_\Delta^\mu$ is the Rarita-Schwinger field of the
$\Delta$-resonance, $\psi_N$ and \boldmath ${\mathbf \pi}$ \unboldmath 
are the nucleon and pion fields, $m_\pi = 0.138$ GeV is the pion mass. 
The coupling constant $f_{\pi N \Delta}$ ($\simeq 2$) will not enter 
explicitly in final expressions for $\Pi(k)$ 
(see Eqs. (\ref{repol}),(\ref{impol}),(\ref{cpol})), therefore, we do not 
need to specify its value. Using Feynman rules and the Lagrangian (\ref{Lagr}) 
the pion polarization function in the case of isospin symmetric nuclear
matter may be written as:
\begin{equation}
      -i\Pi(k) = -{4f_{\pi N \Delta}^2 \over 3m_\pi^2} 
                 \int\, {d^4p \over (2\pi)^4} k_\mu k_\nu
                 \mbox{Sp}\left( G_\Delta^{\mu\nu}(p+k) G_0(p)
                          \right)~,                       \label{Polfun}
\end{equation}
where
\begin{equation}
       G_0(p) = (\not\!p + m_N)\left( {1 \over p^2 - m_N^2 + i\epsilon}
                                  + 2 \pi i\delta(p^2-m_N^2)\theta(p^0)
                                    n({\bf p})
                               \right)                    \label{nucprop}
\end{equation}
is the nucleon propagator in noninteracting nuclear matter 
(c.f. \cite{Dekk94}),
\begin{equation}
       n({\bf p}) = {1 \over \exp((E_N({\bf p})-\mu)/T) + 1}     \label{occnum}
\end{equation}
is the occupation number written for simlicity in the rest frame (r.f.) of 
nuclear matter. 
\begin{equation}
       E_N({\bf p}) = \sqrt{{\bf p}^2+m_N^2}               \label{nfree}
\end{equation}
is the free nucleon energy with $m_N=0.938$ GeV being the nucleon mass. 
For later use we define here also the free $\Delta$ and pion
energies:
\begin{eqnarray}
     E_\Delta({\bf p}) & = & \sqrt{{\bf p}^2+m_\Delta^2}~,  \label{dfree} \\
     E_\pi({\bf k}) & = & \sqrt{{\bf k}^2+m_\pi^2}          \label{pifree}
\end{eqnarray}
with the pole delta mass $m_\Delta=1.232$ GeV. The Rarita-Schwinger 
$\Delta$-resonance propagator $G_\Delta^{\mu\nu}(p)$ is not unambiguously 
defined in the literature. We will use an expression from Ref. \cite{PLM01} 
\begin{equation}
       G_\Delta^{\mu\nu}(p) = -{\not\!p + \sqrt{p^2} \over 
                                 p^2 - m_\Delta^2 
                                + i\sqrt{p^2}\Gamma_\Delta(p^2)}
         \left(g^{\mu\nu} - {1\over3}\gamma^\mu\gamma^\nu
               - {2 p^\mu p^\nu \over 3 p^2} 
               + {p^\mu\gamma^\nu - p^\nu\gamma^\mu \over 3 \sqrt{p^2}}
         \right)~,
                                                    \label{delprop}
\end{equation}
which is similar to the one from Ref. \cite{Dekk94} with an addition of the
term proportional to the total width of the $\Delta$-resonance 
$\Gamma_\Delta(p^2)$ in the denominator and
with a replacement of the pole delta mass $m_\Delta$ by
$\sqrt{p^2}$ in the numerator and in the tensor part.  

The nucleon propagator (\ref{nucprop}) contains the usual vacuum part
$\propto (p^2 - m_N^2 + i\epsilon)^{-1}$ and the in-medium part
$\propto \delta(p^2-m_N^2)$. The latter reflects rescattering of a pion
on real nucleons. The vacuum part contributes to $\mbox{Im}\,\Pi(k)$
only for $k^2 > (m_N+m_\Delta)^2$, which corresponds to the decay of an
off-shell pion to the $\Delta$-resonance and antinucleon. Obviously,
the pions produced in the baryonic resonance decays can not match this
condition. Therefore, we will keep only the in-medium part of the nucleon
propagator. Substituting Eqs. (\ref{nucprop}),(\ref{delprop}) into 
Eq. (\ref{Polfun}) and keeping only the density dependent 
($\propto n({\bf p})$) term in the nucleon propagator we get the following 
expression for the pion polarization function:
\begin{equation}
       \Pi(k) = {16f_{\pi N \Delta}^2 \over 9m_\pi^2}
\int\, {d^3p \over (2\pi)^3 E_N({\bf p})} n({\bf p}) ((pk)^2-m_N^2k^2)
                      {p(p+k)+m_N\sqrt{s} \over
                       (s-m_\Delta^2+i\sqrt{s}\Gamma_\Delta(s))s}~,
                                                           \label{Polfun1}
\end{equation}
where $s=(p+k)^2$ and the nucleon is on its mass shell, i.e.
$p^2 = m_N^2$. In the nonrelativistic limit  $|{\bf k}| \ll m_N$, 
$|{\bf p}| \ll m_N$ and for small $\Delta$ off-shellness 
$|\sqrt{s} - m_\Delta| \ll m_\Delta$ Eq. (\ref{Polfun1}) transforms
to well known formula (c.f. Ref. \cite{EW}):
\begin{equation}
       \Pi(k) \simeq  {16f_{\pi N \Delta}^2 \over 9m_\pi^2} |{\bf k}|^2
\int\, {d^3p \over (2\pi)^3} n({\bf p})
          {1 \over
E_N({\bf p}) + k^0 - E_\Delta({\bf p}+{\bf k}) + 
                      {i \over 2} \Gamma_\Delta(s)}~.        \label{A4}
\end{equation}

In the spirit of the transport theory, however, one would like to express 
the pion polarization function in terms of the $\pi^+ p \to \Delta^{++}$ 
cross section
\begin{equation}
       \sigma_{\pi^+ p \rightarrow \Delta^{++}}(s,k^2) 
    =  {4\pi \over q^2(s,k^2)}
       {2s\Gamma_{\Delta\rightarrow \pi N}(s,k^2)
       \Gamma_\Delta(s) \over
       (s-m_\Delta^2)^2 + s\Gamma_\Delta^2(s)}~,
                                                            \label{sigpind}
\end{equation}
generalized here for the case of the off-shell pion with four-momentum $k$.
\begin{equation}
        q(s,k^2)=\sqrt{ (kp)^2 - k^2 m_N^2 }/\sqrt{s}  \label{qcm}
\end{equation}
is the c.m. momentum of the pion and nucleon while
$\Gamma_{\Delta\rightarrow \pi N}(s,k^2)$ is the decay width of the 
$\Delta$-resonance. The decay width can, in principle, be easily calculated 
within the same Lagrangian of Eq.(\ref{Lagr}) which produces 
the expression (c.f. Ref. \cite{Dekk94}):
\begin{eqnarray}
       \Gamma^{(0)}_{\Delta\rightarrow \pi N}(s,k^2)
 & = & {f_{\pi N \Delta}^2 q(s,k^2) \over 12 \pi m_\pi^2 s^2}
     ((pk)^2-m_N^2k^2)(p(p+k)+m_N\sqrt{s})      \nonumber \\
 & = & {f_{\pi N \Delta}^2 q^3(s,k^2) \over 12 \pi m_\pi^2 \sqrt{s}} 
        ( \sqrt{q^2(s,k^2)+m_N^2} + m_N )~.                  \label{gamdec}
\end{eqnarray}
However, we will use Eq.(\ref{gamdec}) only for the derivation of the 
pion polarization function in terms of the cross section. In numerical
calculations of the pion spectral function we substitute in Eq.(\ref{sigpind})
the off-shell extension of the vacuum width parameterization from 
Ref. \cite{effe1}
\begin{equation}
         \Gamma_{\Delta \rightarrow \pi N}(s,k^2) 
       = \Gamma_0\left({q(s,k^2) \over q_0}\right)^3
         {m_\Delta \over \sqrt{s}} 
         {\beta_0^2+q_0^2 \over \beta_0^2+q^2(s,k^2)} 
                                                             \label{gamdel}
\end{equation}
with $q_0 \equiv q(m_\Delta^2,m_\pi^2)$, $\Gamma_0=0.118$ GeV 
and $\beta_0=0.2$ GeV. The total width of the $\Delta$-resonance in
nuclear medium is taken as
\begin{equation}
       \Gamma_\Delta(s) 
     = \Gamma_{\Delta \rightarrow \pi N}(s,m_\pi^2)~.   \label{gamtot}
\end{equation}

Substituting Eqs.(\ref{sigpind}),(\ref{gamdec}) into Eq.(\ref{Polfun1})
one gets the expression:
\begin{equation} 
       \Pi(k) = {8\over3} \int\, {d^3p \over (2\pi)^3 E_N({\bf p})} n({\bf p})
                   { q(s-m_\Delta^2-i\sqrt{s}\Gamma_\Delta(s))
                     \over \Gamma_\Delta(s) }
                   \sigma_{\pi^+ p \to \Delta^{++}}(s,k^2)~.
                                                          \label{Polfun2}
\end{equation}
Separating now real and imaginary parts we have:
\begin{eqnarray} 
      \mbox{Re}\,\Pi(k) & = & \int\, {2d^3p \over (2\pi)^3}
                                     {m_N \over E_N({\bf p})}
                            n({\bf p}) {p_{rel} (s-m_\Delta^2) \over
                                  \Gamma_\Delta(s) \sqrt{s}}
             {4\over3} \sigma_{\pi^+ p \to \Delta^{++}}(s,k^2)~,
                                                              \label{repol} \\
      \mbox{Im}\,\Pi(k) & = & -\int\, {2d^3p \over (2\pi)^3}
                                      {m_N \over E_N({\bf p})}
                            n({\bf p}) p_{rel}
                 {4\over3} \sigma_{\pi^+ p \to \Delta^{++}}(s,k^2)~,
                                                          \label{impol}
\end{eqnarray} 
where $p_{rel}=q\sqrt{s}/m_N$ is the pion momentum in the r.f.
of a nucleon. By noting that 
$4/3\sigma_{\pi^+ p \to \Delta^{++}} = \sigma_{\pi p \rightarrow \Delta} 
+ \sigma_{\pi n \rightarrow \Delta}$ we see that Eq.(\ref{impol}) can
be simply rewritten in a more compact and intuitive form:
\begin{equation} 
      \mbox{Im}\,\Pi(k) = -{\rho \over 2} 
                        \langle 
                               {m_N \over E_N({\bf p})} p_{rel}
                               ( \sigma_{\pi p \rightarrow \Delta} 
                               + \sigma_{\pi n \rightarrow \Delta} )
                        \rangle~,
                                                          \label{impol1}
\end{equation}
where the averaging is performed with respect to the nucleon momentum
and $\rho$ is the nuclear matter density. Notice, that
Eqs.(\ref{repol}),(\ref{impol}) are explicitly relativistically covariant, 
since the self-energy of a scalar particle is a Lorentz-scalar.

As pointed out in the beginning of this section, in the derivation of the 
pion polarization function we have missed the $u$-diagram. The particular
form (\ref{delprop}) of the spin-3/2 propagator does not permit to 
reliably calculate the $u$-diagram contribution, since the tensor part
has a pole at $p^2=0$ (c.f. \cite{PLM01} for details). However, using the 
usual Rarita-Schwinger propagator (c.f. \cite{Dekk94}) we have checked
that the $u$-diagram contributes mostly in the space-like region $k^2 < 0$,
which is not important for our study, since in BUU only the time-like pions
are propagated.

Finally, we correct the pion polarization function for the repulsive
interaction of holes and $\Delta$-resonances at short distances
(c.f. \cite{HU94}):
\begin{equation}
       \Pi_c(k) = { |{\bf k}|^2 \Pi(k) \over
                    |{\bf k}|^2 - g^\prime \Pi(k) }
                                                          \label{cpol}
\end{equation}
with $g^\prime=0.5$ being the Migdal parameter. Here the pion four-momentum
$k=(k^0,{\bf k})$ has to be taken in the r.f. of nuclear matter. 
In that frame, due to the isotropy of the nucleon Fermi distribution, the 
polarization function $\Pi_c$ depends only on the pion energy and on the 
absolute value of the pion momentum $\Pi_c(k) \equiv \Pi_c(k^0,|{\bf k}|)$.

Fig.~\ref{fig:uopt_vs_k} shows the real and imaginary parts of the pion 
optical potential
\begin{equation}
       U(|{\bf k}|) = {\Pi_c(\varepsilon_k,|{\bf k}|) \over
                       2 E_\pi({\bf k}) }~,                     \label{Uopt}
\end{equation}
where $\varepsilon_k$ is the solution of the dispersion relation 
(c.f. \cite{RW94})
\begin{equation}
       \varepsilon_k^2 = E_\pi^2({\bf k}) 
                       + \mbox{Re}\,\Pi_c(\varepsilon_k,|{\bf k}|)~.
                                                          \label{disprel}
\end{equation}
We see that the real part of the optical potential changes sign 
and the absolute value of the imaginary part reaches its maximum
practically at the same momentum $|{\bf k}| \simeq 0.380$ GeV/c for 
normal nuclear matter case: $\rho=\rho_0$, $T=5$ MeV (solid lines),
where $\rho_0=0.16$ fm$^{-3}$ is the nuclear saturation density. 
This can be qualitatively understood from Eqs.(\ref{repol}),(\ref{impol}) 
since, neglecting for simplicity the Fermi motion, $\mbox{Re}\,\Pi = 0$ and  
$\sigma_{\pi N \rightarrow \Delta}(s)$ is maximal at $s=m_\Delta^2$.
The estimated pion momentum is $|{\bf k}| = 0.300$ GeV/c, which is somewhat 
different from the numerical value above due to the Fermi motion
and the Migdal correction. Increasing temperature (dashed lines) smeares
out the jump of $\mbox{Re}\,U$ and decreases the peak value of
$-\mbox{Im}\,U$ while shifting the peak to lower values of the pion momentum.
Increasing density (dot-dashed lines) acts in the opposite direction. 
These findigs are in an overall agreement with the calculations of 
Ref. \cite{RW94} (see Fig. 2 of this Ref.). An essential difference is
visible only at high momenta $|{\bf k}| > 0.4$ GeV/c where the potentials
of Ref. \cite{RW94} are more repulsive. The pion optical potential is,
however, sensitive to the choice of the nucleon and delta propagators
(relativistic in our work and nonrelativistic in  Ref. \cite{RW94}),
to the presence of the $u$-diagram contribution missed in our calculations
for simplicity and -- most important -- to the choice of the 
$\Delta$-resonance width. In this exploratory work we do not intend to
study in detail all these effects. Fig.~\ref{fig:uopt_vs_k} has a purpose 
to demonstrate only that our simple spectral function calculation is in the 
reasonable agreement with a state-of-art finite temperature calculation of 
Ref. \cite{RW94}.  

Next we determine the pion spectral function 
\begin{equation}
       A_\pi(k) \equiv -{1 \over \pi} \mbox{Im}\,G(k) 
     =   - {1 \over \pi} { \mbox{Im}\,\Pi_c(k) \over
         (k^2 - m_\pi^2 - \mbox{Re}\,\Pi_c(k))^2 
           + (\mbox{Im}\,\Pi_c(k))^2 }                   \label{pispec}
\end{equation}
with $G(k) = (k^2 - m_\pi^2 - \Pi_c(k))^{-1}$ being the pion propagator.
The exact spectral function must satisfy the normalization condition
\begin{equation}
       \int\limits_0^\infty\,d(k^0)^2 A_\pi(k) = 1~.
                                                            \label{norma}
\end{equation}
We checked that the spectral function (\ref{pispec}) satisfies the
condition (\ref{norma}) with the accuracy of 10 \%.

Fig.~\ref{fig:spfun_5mev_rho0} shows the spectral function vs pion invariant 
mass squared $M_\pi^2 \equiv k^2$ and momentum $|{\bf k}|$ for normal 
nuclear matter (density $\rho=\rho_0$ and temperature $T=5$ MeV). There 
are two peaks of $A$ as a function of $M_\pi^2$ at fixed momentum 
$|{\bf k}|$.
At $|{\bf k}| < 0.4$ GeV/c the lower $M_\pi^2$ sharp peak corresponds to 
the pion-like collective mode, while the higher $M_\pi^2$ broad peak
reflects the $\Delta$-hole mode. With increasing momentum the lower
$M_\pi^2$ peak becomes wider and enters the negative $M_\pi^2$
(space-like) region, while the higher $M_\pi^2$ peak gets narrower
and asymptotically approaches the free pion limit 
($M_\pi^2 = m_\pi^2 \simeq 0.02$ GeV$^2$). Therefore, at high momenta
$|{\bf k}| > 0.4$ GeV/c the higher $M_\pi^2$ peak corresponds to the
pion-like mode and the lower $M_\pi^2$ one describes the $\Delta$-hole
excitation. This level crossing is better illustrated in 
Fig.~\ref{fig:omega_vs_k} where along with the peak positions 
of the spectral function in the ($|{\bf k}|,\omega$)-plane 
(full circles) we present also the free pion dispersion relation 
$\omega = E_\pi({\bf k})$ (solid line), the $\Delta$-hole mode
\begin{equation}
       E_{N\Delta}({\bf k}) = 
       \sqrt{   \omega_\Delta({\bf k})
              ( \omega_\Delta({\bf k}) + g^\prime C ) }     \label{Edhole}
\end{equation}
with 
$\omega_\Delta({\bf k}) = E_\Delta({\bf k}) - m_N$ 
and $C = 8/9 (f_{\pi N \Delta}/m_\pi)^2 \rho$ (c.f \cite{HU94})
(dashed line), and the solution $\varepsilon_k$ of the dispersion
relation (\ref{disprel}) (dash-dotted line). 
The peak positions close to the free pion solution are very
well described by the dispersion relation (\ref{disprel}), which has
a characteristic jump at $|{\bf k}| = 0.38$ GeV/c. We also observe,
that broad peaks at higher (lower) energy for small (large) momenta
correspond indeed to the $\Delta$-hole excitation.

The high-momentum structure of the pion spectral function 
($|{\bf k}| > 0.4$ GeV/c) would be modified by taking into account the
higher resonances in the pion polarization function. Therefore, the
pion peak at high momenta is expected to be smeared out by the higher 
resonances. Nevertheless, as shown in Ref. \cite{Teis97} for the
central heavy-ion collisions at energies about 1 A GeV, the higher 
resonance contribution dominates in the pion yield at the c.m. pion kinetic 
energy above 0.6 GeV which corresponds to pion c.m. momentum above 0.7 GeV/c. 
Thus, we expect that the inclusion of higher baryonic resonances in the pion 
spectral function will not strongly modify the results of our work 
(c.f. Fig.~\ref{fig:piptsp_au106au_new} later). 

In order to estimate an effect of the in-medium modifications of the
$\Delta$-resonance width, we replaced $\Gamma_\Delta(s)$ in Eqs. 
(\ref{sigpind}),(\ref{repol}) by $\Gamma_\Delta(s)(1 - \bar n) + 
\Gamma_{sp}$, where $\Gamma_{sp} = -2 \mbox{Im}\,W_{sp} = 80 \rho/\rho_0$ 
[MeV] is the phenomenological spreading width given by the imaginary part 
of the $\Delta$-spreading potential \cite{Hir79}. $\bar n$ is the angle
averaged nucleon occupation number. The $\Delta$-spreading potential 
takes into account the quasielastic scattering, two- and three-body 
absorption contributions \cite{Oset87,Ko89}. In Fig.~\ref{fig:spfun_spr} we 
compare the spectral functions at $\rho = 2\rho_0$ and $T=70$ MeV calculated 
with vacuum (dashed lines) and in-medium (solid lines) $\Delta$-resonance
width. We see that the $\Delta$-spreading potential smeares-out the
peaks of the spectral function. There is a tendency of the strength
concentration near $M_\pi^2 = m_\pi^2 = 0.02$ GeV$^2$. Correspondingly,
the strength at negative and large positive $M_\pi^2$ is reduced somewhat
by in-medium $\Delta$-width.
 
\section{ Kinetic equation for off-shell pions }

For the description of the off-shell particle dynamics we will use
the spectral distribution functions (c.f. \cite{EM99})
$F_\alpha({\bf r},t,{\bf p},M^2)$ with $\alpha$ being the particle 
type ($\alpha = N, R, \pi$ for nucleons, resonances and pions
respectively). The number of particles of type $\alpha$ in the phase
space element $d^3r d^3p dM^2/(2\pi)^3$ is $F_\alpha d^3r d^3p dM^2/(2\pi)^3$.
The following relation holds:
\begin{equation}
       F_\alpha({\bf r},t,{\bf p},M^2) 
     = f_\alpha({\bf r},t,{\bf p},M^2) 
       A_\alpha({\bf r},t,{\bf p},M^2)                  \label{F}
\end{equation}
where $f_\alpha({\bf r},t,{\bf p},M^2)$ is the usual phase space density,
and $A_\alpha({\bf r},t,{\bf p},M^2)$ is the spectral function. 
The pion spectral function has been defined earlier by Eq.(\ref{pispec}).
The spectral function of a baryonic resonance is taken as the vacuum one:
\begin{equation}
       A_R(M^2) = {1 \over \pi} {M\Gamma_R(M^2) 
                                   \over
                  (M^2-m_R^2)^2 + M^2\Gamma_R^2(M^2)}
                                                        \label{Rspec}
\end{equation}
with $m_R$ being the pole mass and $\Gamma_R^2(M^2)$ being the vacuum
width of the resonance \cite{effe1}. The effect of the nucleon 
off-shellness on the pion and kaon production has been studied 
in Ref. \cite{EM99}. In the present work we treat, however, nucleons
on-shell in order to underline the genuine pion off-shellness effects
on the observables. Thus, we use
\begin{equation}
        A_N(M^2) = \delta(M^2-m_N^2)                    \label{Nspec}
\end{equation}
and everywhere below the nucleon four-momentum is 
$p_N = (E_N({\bf p}_N),{\bf p}_N)$.

It is instructive to write down explicitly the coupled transport equations
for resonances and pions:
\begin{eqnarray}
    \left( 
           {\partial \over \partial t}
       +   {\partial H_{mf}^R \over \partial {\bf p}_R}
           {\partial \over \partial {\bf r}}
       -   {\partial H_{mf}^R \over \partial {\bf r}}
           {\partial \over \partial {\bf p_R}}
    \right) 
           F_R({\bf r},t,{\bf p}_R,M_R^2)  & = & I_R~,    \label{kineqR} \\
    \left( 
           {\partial \over \partial t} 
       +   {{\bf k} \over k^0}
           {\partial \over \partial {\bf r}}
    \right) 
           F_\pi({\bf r},t,{\bf k},M_\pi^2) 
       +   L_{off-shell}[F_\pi]            & = & I_\pi~,    \label{kineqpi}
\end{eqnarray}
where $H_{mf}^R({\bf r},t,{\bf p}_R)$ is the mean field Hamiltonian 
acting on resonances: in this work we take it the same as for nucleons.
$I_R$ and $I_\pi$ are the collision integrals. The term 
$L_{off-shell}[F_\pi]$ on the l.h.s. of (\ref{kineqpi}), which will be
specified later on, is introduced in order to bring pions back on the mass 
shell once they reach the vacuum. 
The collision integrals are expressed in terms
of gain $(<)$ and loss $(>)$ terms as follows:
\begin{eqnarray}
        \lefteqn{ I_R = I_R^< - I_R^> 
           =   \sum_{N,\pi}\,{1 \over 2p_R^0}
                \int\,{g_N d^3p_N \over (2\pi)^3 2E_N({\bf p}_N)}
                \int\,{d^3k dM_\pi^2 \over (2\pi)^3 2k^0}    
                }
       &  &                                       \nonumber \\
       &\times&         
                (2\pi)^4 \delta^{(4)}(p_R-k-p_N) 
                |{\cal M}_{R \to N\pi}|^2         \nonumber \\
       &\times&         
                \{
                  F_\pi({\bf k},M_\pi^2)f_N({\bf p}_N)
                  A_R(M_R^2)(1-f_R({\bf p}_R,M_R^2)) \nonumber \\
         &-&      F_R({\bf p}_R,M_R^2)A_\pi({\bf k},M_\pi^2)
                  (1+f_\pi({\bf k},M_\pi^2))(1-f_N({\bf p}_N))
                \}~,
                                                          \label{cintR} \\
       \lefteqn{ I_\pi = I_\pi^< - I_\pi^> 
             =  \sum_{R,N}\,{1 \over 2k^0}
                  \int\,{g_R d^3p_R dM_R^2 \over (2\pi)^3 2p_R^0}
                  \int\,{g_N d^3p_N \over (2\pi)^3 2E_N({\bf p}_N)}        
               }
       &   &                                              \nonumber \\
       &\times&
                (2\pi)^4 \delta^{(4)}(p_R-k-p_N)
                |{\cal M}_{R \to N\pi}|^2                 \nonumber \\
       & \times &
                  \{
                    F_R({\bf p}_R,M_R^2)A_\pi({\bf k},M_\pi^2)
                    (1+f_\pi({\bf k},M_\pi^2))(1-f_N({\bf p}_N)) \nonumber \\
           &-&    F_\pi({\bf k},M_\pi^2)f_N({\bf p}_N)   
                    A_R(M_R^2)(1-f_R({\bf p}_R,M_R^2))
                  \}~,
                                                           \label{cintpi}
\end{eqnarray}
where we dropped space and time variables for brevity.
In Eqs.(\ref{cintR}),(\ref{cintpi}) $|{\cal M}_{R \to N\pi}|^2$ is the 
invariant matrix element squared averaged over spins of initial and final
particles, $g_\alpha = 2s_\alpha + 1$ ($\alpha = N,R$) is the spin degeneracy
factor. For simplicity, we did not display the contributions of
$N N \leftrightarrow N R$ processes to $I_R$ and of the direct processes 
$N N \leftrightarrow N N \pi$ to $I_\pi$. However, these processes are 
always included in BUU. Sums in Eqs.(\ref{cintR}),(\ref{cintpi}) run 
over all possible isospin states of the particles. Noting that the 
resonance decay width to a nucleon and an off-shell pion of mass $M_\pi$ 
in the resonance r.f. is given by the expression
\begin{eqnarray}
       \lefteqn{ \Gamma_{R \to N \pi}(M_R^2,M_\pi^2)
                 = {1 \over 2M_R}
                 \int\,{g_N d^3p_N \over (2\pi)^3 2E_N({\bf p}_N)}
                 \int\,{d^3k \over (2\pi)^3 2k^0} }
      &  &                                                 \nonumber \\    
      &\times&
                (2\pi)^4 \delta^{(4)}(p_R-k-p_N)
                |{\cal M}_{R \to N\pi}|^2                   
        =  \int\,d\Omega_{c.m.} 
            {1 \over 32 \pi^2 M_R^2} 
            |{\cal M}_{R \to N\pi}|^2 q g_N
                                                            \label{gammar}
\end{eqnarray}
and using the fact that the spin-averaged matrix element does not depend
on the center-of-mass scattering angles, one can rewrite Eq.(\ref{cintR})
as follows:
\begin{eqnarray}
        I_R & = &  \sum_{N,\pi} \int\,dM_\pi^2 {M_R \over p_R^0}
                   \Gamma_{R \to N \pi}(M_R^2,M_\pi^2)         \nonumber \\
       & \times &  \int\,{d\Omega_{c.m.} \over 4\pi}
                   \{ 
                     F_\pi({\bf k},M_\pi^2)f_N({\bf p}_N)
                     A_R(M_R^2)(1-f_R({\bf p}_R,M_R^2)) \nonumber \\
            & - &    F_R({\bf p}_R,M_R^2)A_\pi({\bf k},M_\pi^2)
                  (1+f_\pi({\bf k},M_\pi^2))(1-f_N({\bf p}_N))
                   \}~.                                      \label{cintR1}
\end{eqnarray}
Introducing the cross section 
\begin{equation}
        \sigma_{\pi N \to R}(M_R^2,M_\pi^2)
      = {4\pi^2 \over q^2} \Gamma_{R \to N \pi}(M_R^2,M_\pi^2)
        M_R A_R(M_R^2) {g_R \over g_N}                       \label{sigpinr}
\end{equation}
the pion collision integral can be also simplified as
\begin{eqnarray}
        I_\pi & = & \sum_{R,N} \int\,{g_N d^3p_N \over (2\pi)^3}
                   v_{\pi N} \sigma_{\pi N \to R}(M_R^2,M_\pi^2) \nonumber \\
         & \times & 
                    \{
                    f_R({\bf p}_R,M_R^2)A_\pi({\bf k},M_\pi^2)
                    (1+f_\pi({\bf k},M_\pi^2))(1-f_N({\bf p}_N)) \nonumber \\
           &-&        F_\pi({\bf k},M_\pi^2)f_N({\bf p}_N)   
                      (1-f_R({\bf p}_R,M_R^2))
                    \}
                                                             \label{cintpi1}
\end{eqnarray}
with $v_{\pi N} = \sqrt{(kp_N)^2-M_\pi^2m_N^2}/(k^0E_N({\bf p}_N))$ being 
the relative velocity of the pion and nucleon.

In the following we neglect for simplicity the second-order effect of the 
pion off-shellness on the resonance width. Thus, we replace in 
Eq.(\ref{cintR1}) and, to maintain detailed balance, also in Eq.(\ref{sigpinr})
\begin{equation}
       \Gamma_{R \to N \pi}(M_R^2,M_\pi^2)
                 \rightarrow 
       \Gamma_{R \to N \pi}(M_R^2,m_\pi^2)~,                 \label{replace}
\end{equation}
i.e. the off-shellness of the pion in the resonance decay vertex 
is neglected.
We have to comment that this simplification has been done only in the BUU
implementation. (In the pion polarization function calculation the full
off-shell decay width (\ref{gamdel}) has been applied.) Correspondingly,
the mass squared of the outgoing pion in the resonance decay has been sampled
in the interval $[0;(M_R-m_N)^2]$ according to the distribution
$\propto A_\pi({\bf k},M_\pi^2)$. In the case of the
direct (background) pion production $N N \to N N \pi$ the value of
$M_\pi^2$ has been sampled in the interval $[0;(\sqrt{s}-2m_N)^2]$
using the same distribution.

The propagation of the off-shell pion between its production
in the resonance decay (or in the direct process $N N \to N N \pi$)
and absorption in the collision with a nucleon $\pi N \to R$ 
(or with two nucleons $\pi N N \to N N$) is described by the l.h.s.
of Eq.(\ref{kineqpi}). In BUU the test particle method is used to 
solve the transport equations. The pion spectral distribution function
is represented by test particles in the following way:
\begin{equation}
       F_\pi({\bf r},t,{\bf k},M_\pi^2)
   = \sum_i\,\delta({\bf r}-{\bf r}_i(t))
             \delta({\bf k}-{\bf k}_i(t))
             \delta(M_\pi^2-M_{\pi,\,i}^2(t))~.          \label{testpart}
\end{equation}
Following Refs. \cite{effe1,EM99} we introduce a scalar off-shell potential 
$s_{\pi,\,i}$ acting on the i-th pion test particle:
\begin{equation}
       s_{\pi,\,i}({\bf r}_i(t),t) 
    = { M_{\pi,\,i}(t_{cr}) - m_\pi 
        \over \rho_N({\bf r}_i(t_{cr}),t_{cr}) }
      \rho_N({\bf r}_i(t),t)~,                           \label{fakepot}
\end{equation}
where $\rho_N({\bf r},t)$ is the density of nucleons in the local r.f.
of nuclear matter. $t_{cr}$ is the production time of the test particle.
The potential (\ref{fakepot}) drives the off-shell pion back to its
mass shell when it leaves the nucleus, i.e. we assume that
\begin{equation}
       M_{\pi,\,i}(t) = m_\pi + s_{\pi,\,i}({\bf r}_i(t),t)~.
                                                         \label{pimass}
\end{equation}
Therefore, Eqs.(\ref{fakepot}),{\ref{pimass}) provide the correct
boundary condition for large times, i.e. when pions reach the vacuum:
\begin{equation}
       \lim_{t \to \infty}\, M_{\pi,\,i}(t) = m_\pi~.          \label{limit}
\end{equation}
The operator $L_{off-shell}[F_\pi]$ in the l.h.s. of Eq.(\ref{kineqpi})
is defined by its action on the distribution function in the test 
particle representation (\ref{testpart}) as follows:
\begin{eqnarray}
       L_{off-shell}[F_\pi] 
 & \equiv & \sum_i\, 
              \left(
                    2 M_{\pi,\,i}(t) \dot{s}_{\pi,\,i} 
                   {\partial \over \partial M_\pi^2}
              - {\partial H_{off-shell,\,i} \over \partial {\bf r}_i}
                {\partial \over \partial {\bf k}}
              \right)                                    \nonumber \\
 & \times & \delta({\bf r}-{\bf r}_i(t))
             \delta({\bf k}-{\bf k}_i(t))
             \delta(M_\pi^2-M_{\pi,\,i}^2(t))~,            \label{Los}
\end{eqnarray}
where
\begin{equation}
       H_{off-shell,\,i} = \sqrt{ {\bf k}_i^2 + 
                                (m_\pi + s_{\pi,\,i})^2 }       \label{Hos}
\end{equation}
is the energy of the pion test particle $i$. From Eqs.(\ref{kineqpi}),
(\ref{Los}) we get equations of motion of test particles in coordinate
and momentum space:
\begin{eqnarray}
       \dot{\bf r}_i & = & {\partial H_{off-shell,\,i} 
                      \over \partial {\bf k}_i}~,          \label{Hameq1} \\
       \dot{\bf k}_i & = & -{\partial H_{off-shell,\,i} 
                      \over \partial {\bf r}_i}~.          \label{Hameq2}
\end{eqnarray}
These two Hamiltonian equations together with Eq.(\ref{pimass}) determine
the evolution of the spectral distribution function (\ref{testpart})
completely.

Due to the explicit time dependence of the potential (\ref{fakepot}),
which acts as an external field, the energy of the whole system is not
conserved. However, in practice, the violation of the energy conservation
due to the off-shell potential turns out to be very small ($\sim 0.2$ 
MeV/nucleon for the central Au+Au collision at 1 A GeV, i.e.  
$\sim 0.1\%$ of the total c.m. kinetic energy per nucleon). This is caused 
by a relatively small number of pions with respect to the baryon number.

Some comments are in order on the relation of the phenomenological approach
of Refs. \cite{effe1,EM99} applied in the present work
with the more general test particle equations of motion derived in Refs. 
\cite{CaJu00,Leu00} on the basis of the Kadanoff-Baym equations. 
Indeed, Eqs.(\ref{fakepot}),(\ref{pimass}) which determine the evolution
of the pion mass and the Hamiltonian equations (\ref{Hameq1}),(\ref{Hameq2})
can be obtained from the formalism of Refs. \cite{CaJu00,Leu00} by
assuming that $\mbox{Im}\,\Pi_c \propto \rho_N$ and neglecting
$\mbox{Re}\,\Pi_c$ in the propagation of the pions. 
Earlier BUU studies \cite{Helg98,Xiong93,Fuchs97} within the on-shell 
approximation have shown that the inclusion of the real part of the pion 
optical potential results in an enhanced soft pion yield.
However, our attempted treatment 
of the off-shell pion propagation within the formalism of Refs. 
\cite{CaJu00,Leu00} with the $\Delta$-hole pion polarization function 
resulted in the appearance of the superluminous pions. We have, therefore, 
chosen the simplified Hamiltonian approach of Refs. \cite{effe1,EM99} for 
the propagation of the off-shell pions which has no this problem, since the 
off-shell potential $s_{\pi,\,i}$ is momentum-independent.

\section{ Numerical results }

The pion spectral function has been implemented in the BUU model in
the version of Ref. \cite{effe1}. The temperature $T$ and the baryon
density $\rho$ have been calculated locally in BUU and their values
have been used to interpolate $A_\pi$ which was stored on the 
4-dimensional grid in the space $(M_\pi^2,|{\bf k}|,\rho,T)$.
This allows one to save CPU time by avoiding on-line calculation of $A_\pi$. 
Masses of outgoing pions from resonance
decays or from NN collisions have been sampled according to $A_\pi$.
The pions have then been propagated by using Eqs.(\ref{Hameq1}),(\ref{Hameq2})
until they have been reabsorbed in $\pi N \to R$ or $\pi N N \to N N$ 
processes or emitted to vacuum.
The BUU calculations of the present work are performed using the soft 
momentum-dependent mean field (SM) with the incompressibility $K=220$ MeV
and employing the quenching of the resonance production in order to get
the correct pion multiplicity \cite{LCLM01}. Below, if the opposite 
is not stated explicitly, the vacuum $\Delta$-width (\ref{gamdel})
is used in the pion spectral function. 

Fig.~\ref{fig:dm_evol} shows the time evolution of the pion mass 
distribution in the central collision Au+Au at 1.06 A GeV. At early times 
($t=10 \div 20$ fm/c) a rather broad range in $M_\pi$ is populated by the 
resonance decays in dense nuclear matter, where the pion spectral function 
is broad. 
In order to separate the effect of the off-shell transport by 
Eqs.(\ref{Hameq1}),(\ref{Hameq2}) we performed also the calculation 
putting $L_{off-shell}[F_\pi] = 0$ in Eq.(\ref{kineqpi}) (dash-dotted 
lines). In this case pions emitted to vacuum do not lose their off-shellness
and the $M_\pi$-spectrum stays quite broad even at $t=40$ fm/c when all
violent processes are already finished and the system is already quite
dilute (central baryon density $\simeq 0.02$ fm$^{-3}$). The full 
calculation (solid lines) results in a strong narrowing of the 
$M_\pi$-spectrum at the late stage of the reaction: at $t=40$ fm/c 
the most part of pions is already on the mass shell. Continuing the
calculation until 80 fm/c (dotted line on lower right panel) produces
a rather moderate variation of the $M_\pi$-spectrum towards precise
$\delta(M_\pi-m_\pi)$. Thus, for the comparison with experimental
data below we have stopped calculations at $t=40$ fm/c.

In Fig.~\ref{fig:piptsp_au106au_new} we present the inclusive transverse 
momentum spectra at midrapidity for the same system Au+Au at 1.06 A GeV in 
comparison with the experimental data from Refs. \cite{Pelte97,Schwalb94}. 
The calculation with on-shell pions (dashed lines) describes reasonably well 
the high-$p_t$ part of the spectra, but underpredicts the data at low $p_t$. 
Taking into account the pion off-shellness (dotted lines) increases the 
low-$p_t$ pion yields improving, thus, the agreement with the data. The 
low-$p_t$ pion enhancement is caused by the off-shell pion propagation. 
To check this, we switched off again the $L_{off-shell}[F_\pi]$ term 
in Eq.(\ref{kineqpi}) (dash-dotted lines).
We see that, indeed without this term the low-$p_t$ enhancement disappears.
The effect is caused by the low mass pions ($M_\pi < m_\pi$) produced by the 
resonance decays at early stage of the collision (Fig.~\ref{fig:dm_evol}).
These pions experience the attractive off-shell potential (\ref{fakepot}),
and, therefore, they are decelerated propagating out of the dense region of 
the nuclear system.

One may notice also in Fig.~\ref{fig:piptsp_au106au_new} a slightly enhanced 
yield of high-$p_t$ pions by the off-shell calculation with vacuum 
$\Delta$-width. The agreement with data at large transverse momenta 
can be improved by using the $\Delta$-spreading width (solid lines). 
This is caused by the reduced strength of the pion spectral 
function at large pion masses (c.f. Fig.~\ref{fig:spfun_spr}).

\section{ Summary and outlook }

In this work we have generalized the BUU approach for the pion
off-shellness in the nuclear medium. The $\Delta$-hole model
has been applied to calculate the pion spectral function which has
been used, then, in BUU to choose randomly the outgoing pion mass in the
resonance decay. The propagation of the produced off-shell
pions has been described in a Hamiltonian approach -- developed earlier
in Refs. \cite{effe1,EM99} -- using the baryon density dependent 
off-shell scalar potential which acts on pions.
This guarantees that a pion emitted to vacuum is always on the mass shell.

As an observable effect in heavy-ion collisions at 1 A GeV, we have 
demonstrated the low $p_t$ enhancement of the pion yield due to
the off-shell pion propagation, which is in agreement with earlier
BUU calculations \cite{Helg98,Xiong93,Fuchs97} using the in-medium 
pion dispersion relation.
We expect, that the pion off-shellness will also influence other
observables, which are not covered in this work, e.g. the $K^+$, $K^-$
and $\rho$ production via reactions $\pi N \to K^+ \Lambda$,
$\pi \Lambda \to K^- N$ and $\pi\pi \to \rho$ respectively (study in
progress).

Another important aspect is the modification of the $\Delta$-width due
to coupling to an off-shell pion. We have studied this effect assuming
the in-medium broadening of the $\Delta$-resonance given by the 
phenomenological $\Delta$-spreading potential \cite{Hir79}. Overall, a 
rather weak influence of the in-medium $\Delta$-width on pion
observables is found. We have to remark, however, that we included the 
in-medium $\Delta$-width in the pion spectral function only, not in
BUU. The explicit in-medium modification of the $\Delta$-width in BUU has 
been studied earlier in Refs. \cite{Helg98,EHTM97}, where a quite weak 
effect on pion observables in heavy-ion collisions and on 
$\gamma$-absorption cross section on nuclei was found. 
Finally, it is straightforward to generalize the $\Delta$-hole 
model including higher baryonic resonances which will then also contribute 
to the pion spectral function.

\begin{acknowledgments}
Stimulating discussions with W. Cassing, S. Leupold, M. Post and V. Shklyar 
are gratefully acknowledged. The authors are especially grateful to
Stefan Leupold for the careful reading of the manuscript, useful advices
and comments. 
\end{acknowledgments}

\clearpage

\thispagestyle{empty}

\begin{figure}

\vspace{5cm}

\includegraphics{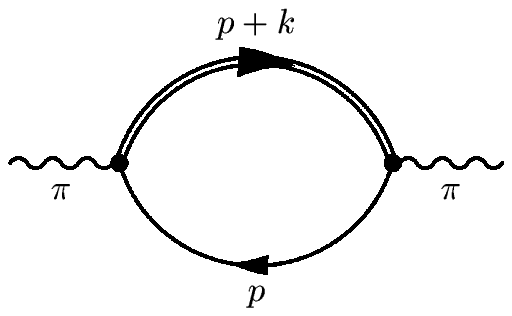}

\vspace{7cm}

\caption{\label{fig:dhole} Feynman graph representing $-i\Pi(k)$
with $\Pi(k)$ being the pion polarization function. $k$ and $p$ are
the pion and nucleon four-momenta respectively.}
\end{figure}

\clearpage

\thispagestyle{empty}

\begin{figure}

\vspace{8cm}

\includegraphics{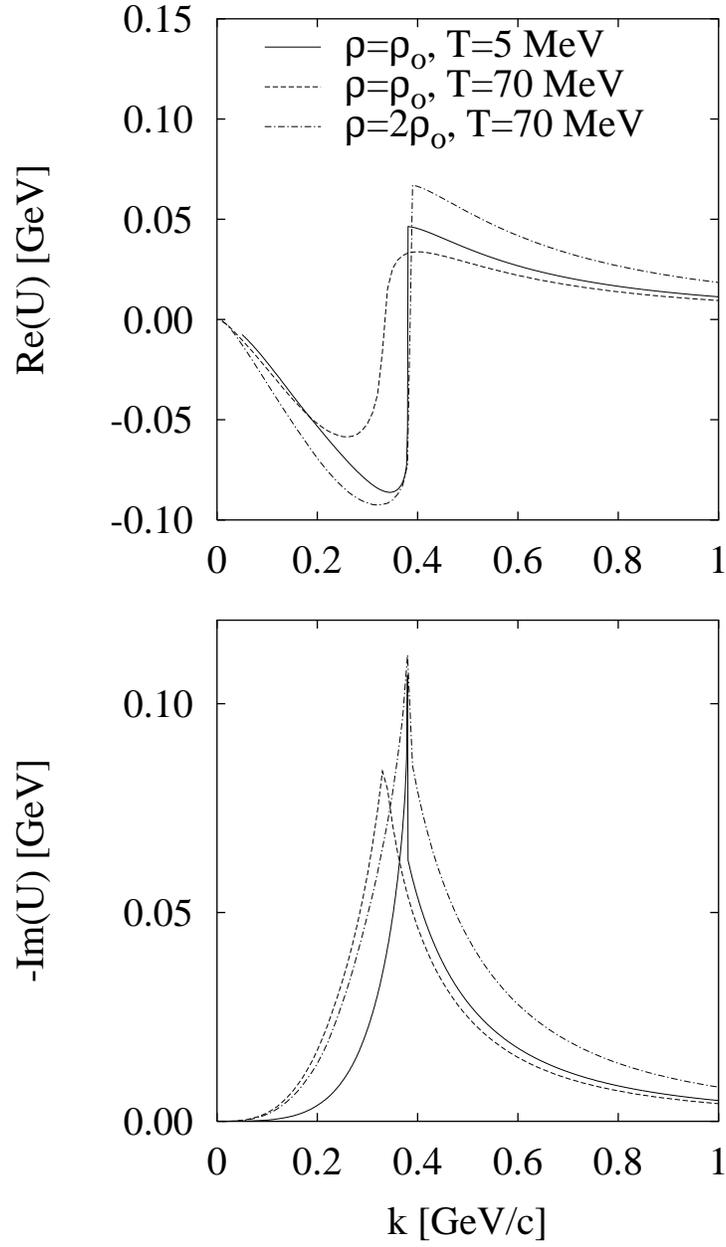}

\vspace{4cm}

\caption{\label{fig:uopt_vs_k} Real (upper panel) and imaginary
(lower panel) parts of the pion optical potential as functions of the 
pion momentum at various densities and temperatures.}
\end{figure}

\clearpage

\thispagestyle{empty}

\begin{figure}

\vspace{4cm}

\includegraphics{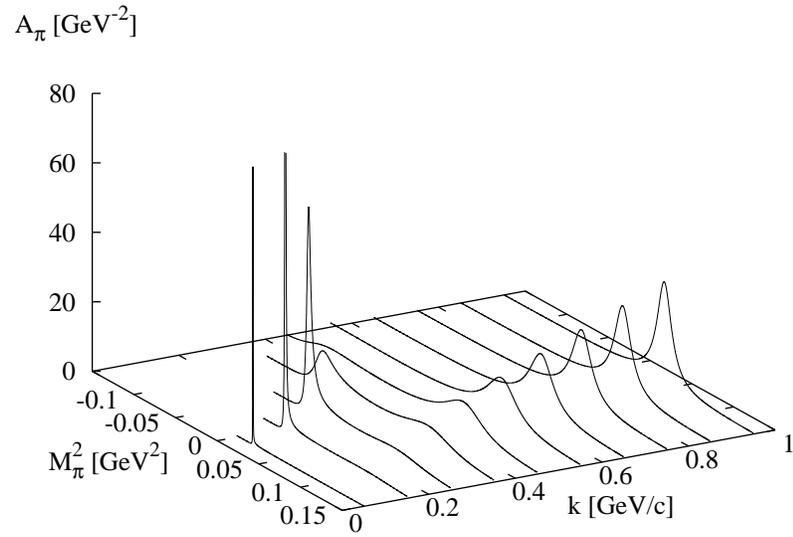}

\vspace{5cm}

\caption{\label{fig:spfun_5mev_rho0} Pion spectral function in nuclear
matter at the density $\rho=\rho_0$ and temperature $T=5$ MeV shown
vs mass squared and momentum.}
\end{figure}

\clearpage

\thispagestyle{empty}

\begin{figure}

\vspace{4cm}

\includegraphics{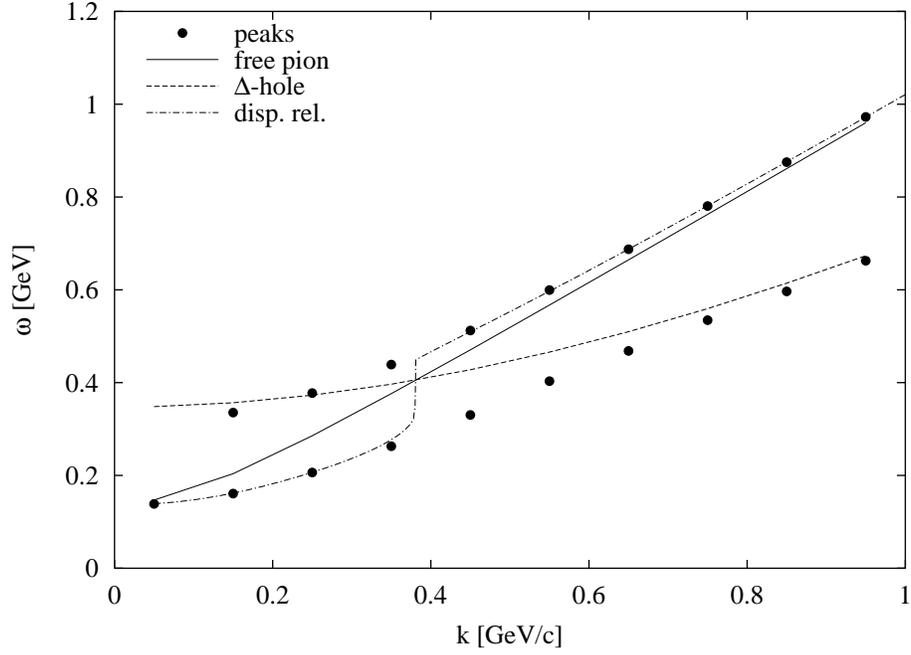}

\vspace{5cm}

\caption{\label{fig:omega_vs_k} Peak positions of the pion spectral
function $A_\pi$ in the ($|{\bf k}|,\omega$)-plane for nuclear matter at 
the density $\rho=\rho_0$ and temperature $T=5$ MeV -- full circles.
Free pion dispersion relation $E_\pi({\bf k})$ is shown by the solid line.
Dashed line shows the $\Delta$-hole energy $E_{N\Delta}({\bf k})$.
Solution $\varepsilon_k$ of the dispersion relation (\ref{disprel}) 
is depicted by dash-dotted line.}
\end{figure}

\clearpage

\thispagestyle{empty}

\begin{figure}

\vspace{8cm}

\includegraphics{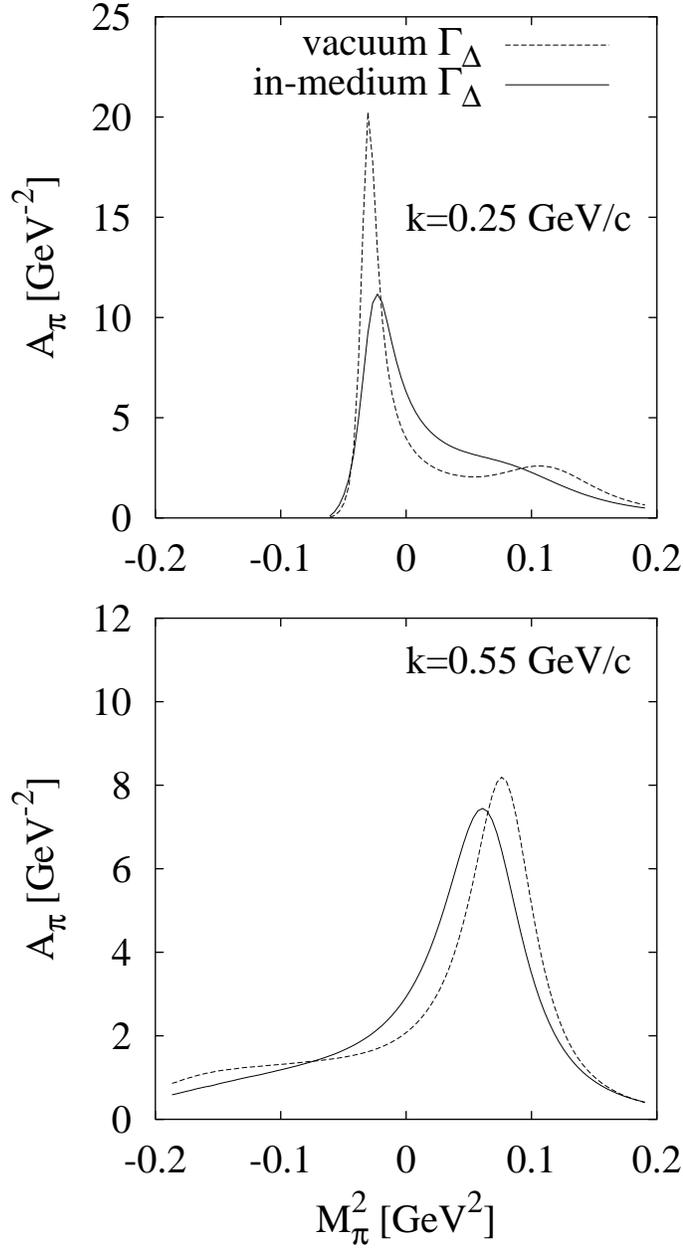}

\vspace{4cm}

\caption{\label{fig:spfun_spr} Pion spectral function vs pion invariant
mass squared at fixed pion momentum $|{\bf k}|=0.25$ GeV/c (upper panel)
and $|{\bf k}|=0.55$ GeV/c (lower panel) in nuclear matter at 
$\rho=2\rho_0$ and $T=70$ MeV. Dashed and solid lines present the 
calculations with vacuum and in-medium $\Delta$-width respectively.}
\end{figure}

\clearpage

\thispagestyle{empty}

\begin{figure}

\vspace{1cm}

\includegraphics{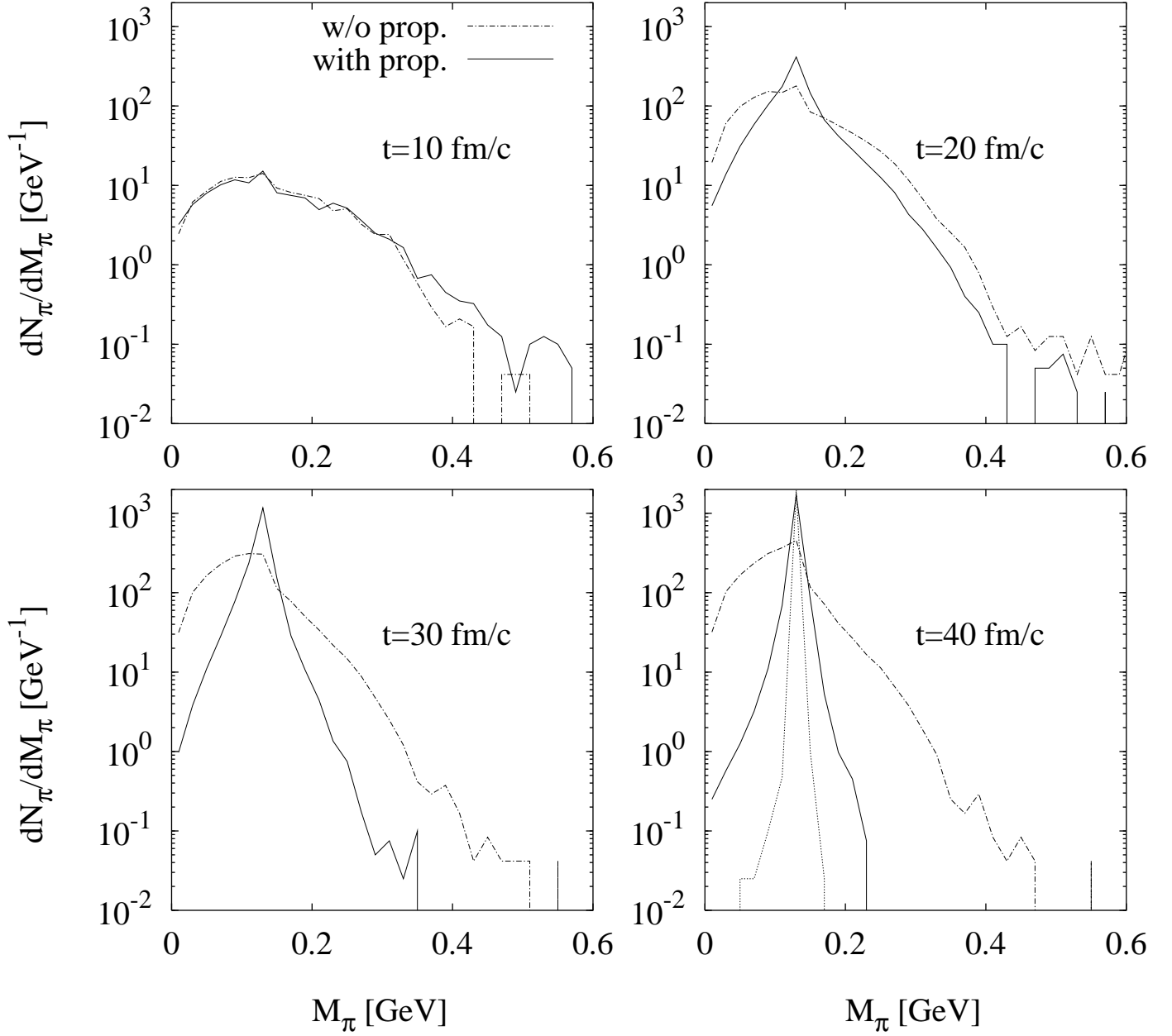}

\vspace{1cm}

\caption{\label{fig:dm_evol} Time evolution of the pion mass distribution
in the central $Au+Au$ collision at 1.06 A GeV. 
Solid and dash-dotted lines show the results of calculations with and
without the off-shell term $L_{off-shell}[F_\pi]$ in pion propagation 
respectively. Dotted line in the lower right panel shows the mass 
distribution at $t=80$ fm/c for the calculation with off-shell term.}
\end{figure}

\clearpage

\thispagestyle{empty}

\begin{figure}

\vspace{1cm}

\includegraphics{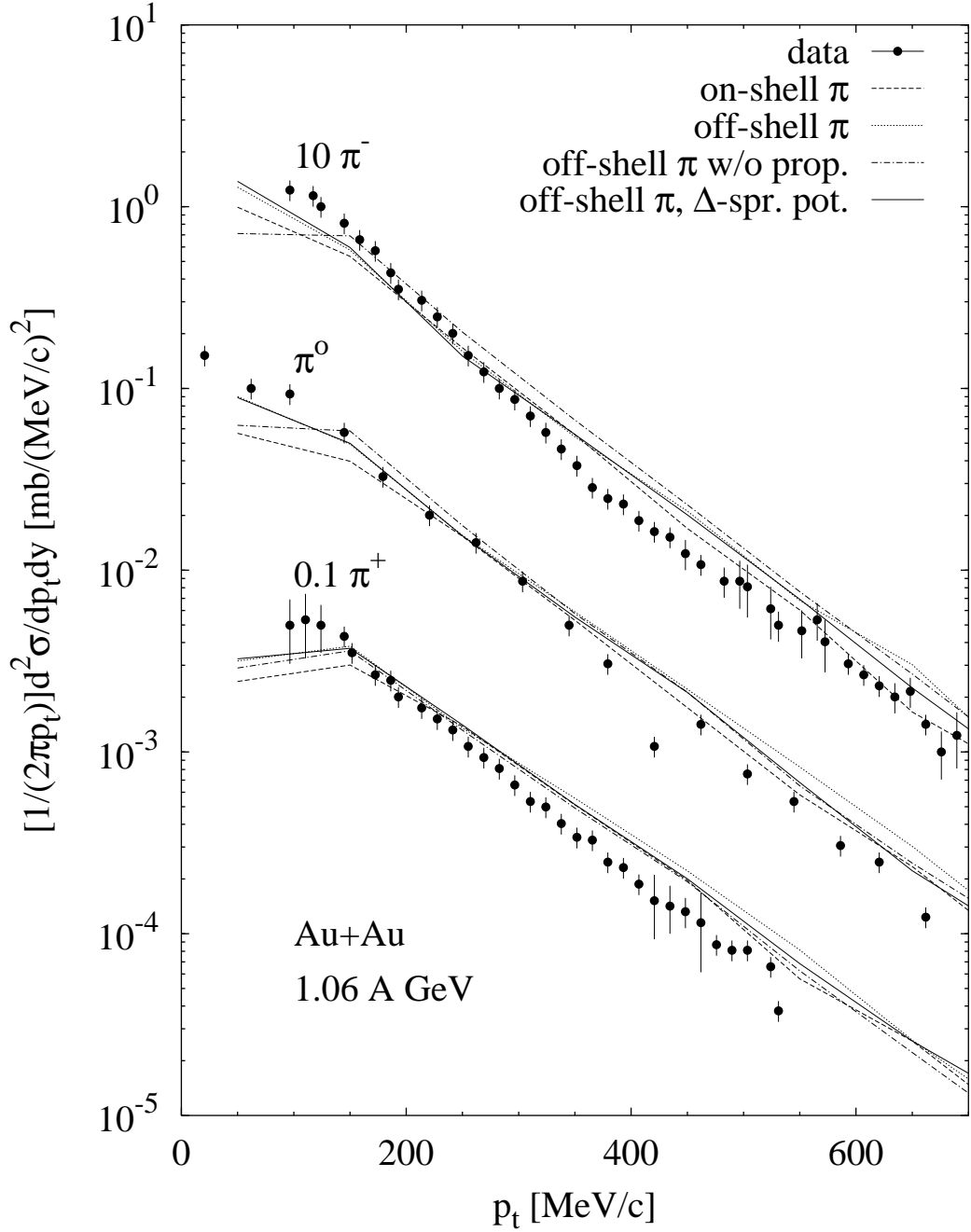}

\vspace{1cm}

\caption{\label{fig:piptsp_au106au_new} Pion transverse momentum spectra
at midrapidity for Au+Au collisions at 1.06 A GeV in comparison with
experimental data from Refs. \cite{Pelte97,Schwalb94}. Dashed lines 
show the calculation with on-shell pions. Dotted and dash-dotted lines
present the off-shell calculations using vacuum $\Delta$-width with and 
without $L_{off-shell}[F_\pi]$ term respectively. Solid lines show
the off-shell calculation with in-medium $\Delta$-width.}
\end{figure}

\end{document}